UNIVERSAL BEHAVIOR OF A RESEARCH PRODUCTIVITY INDEX

P. D. Batista, M. G. Campiteli, O. Kinouchi and A. S. Martinez

Faculdade de Filosofia, Ciências e Letras de Ribeirão Preto

Universidade de São Paulo, Brazil

Recently Hirsch [1,2] has proposed a new scalar index $h$ to quantify individual's scientific research impact. A researcher with index $h$ has $h$ papers with at least $h$ citations. This index has several advantages: (i) it combines productivity with impact, (ii) the necessary data is easy to access at the Thomson ISI Web of Science database, (iii) it is not sensitive to extreme values, (iv) it is hard to inflate. However, this index remains sensitive to the research field. It is even difficult to compare researchers from different areas within a given discipline. Further, since $h$ is an integer number, many researchers may have the same index $h$, so that discriminating or listing them demands further indexes.

The number of authors correlates positively with the number of citations that a paper receives in a given time interval [3]. It is also reasonable to assume that the co-authorship behavior is characteristic of each discipline. Here we propose a new index $h_I$ which deals with these issues in a robust way [4]. The statistics of $h$ and $h_I$ are presented for the fundamental research fields in Brazil. The relatively small size of this community enabled us to do a complete index statistics.

From Thomson ISI Web of Science, we have compiled a database of the Brazilian scientific research in four different fields: Physics, Chemistry, Biology/Biomedics and Mathematics. The research has been conducted using the query "Brazil OR Brasil" in the address field, considering all documents published from 1945 up to 2004. Our database contains information of about 188,909 bibliographical references (additional information is available in the supplementary material).

To account for co-authorship, divide $h$ by the mean number of researchers in the $h$ publications: $<N_a> = N_a^{(T)} / h$ where $N_a^{(T)}$ is the total number of authors (author multiple occurrences are allowed) in the considered $h$ papers, leading to:

$$h_I = h/\langle N_a \rangle = h^2/N_a^{(T)} \qquad (1)$$

which gives additional information about the research relevance. The rationale for this procedure is that we want to measure the average individual productivity. Further, more authors could produce more future self-citations which may inflate the statistics. Once $h$ has been computed, the $h_I$ index is also easy to be obtained. The rank plots of $h_I$ and $h$ (Fig. 1-A and inset, respectively) for the brazilian scientific community present different behavior. In contrast to the $h$ curves, the $h_I$ curves are smoother and have the same functional shape for all disciplines. We have found that biologists have smaller $h$ than physicists in Brazil, contrasting to Hirsch's observations about worldwide data [1]. This may be due to the lack of financial support to sustain the experimental nature of Biology, in Brazil.

We have also calculated the $h$ and $h_I$ for the top ten physicists and biologists listed in [1]. Contrasting to the $h$ averages, the average of $h_I$ for both areas are similar and the maximum values of $h_I$ are closer than $h$ maximum values (data available on the supplementary information).

The index $h_I$ is complementary to $h$. It lifts the $h$ degenerescency and has the advantage of being less sensitive to different research fields. This allows a less biased comparison due to the consideration of co-authorship. However, the use of the mean in $h_I$ index could penalize authors with eventual papers with large co-authorship. A possible correction to this factor (a little bit harder to obtain) is to consider the median instead of the mean value.

The fundamental result of this communication is that there is a universal behavior for the relative index $h_I/\langle h_I \rangle$ as a function of the relative rank $R/R_{max}$ as shown in Fig.1-b, where $\langle h_I \rangle$ and rank maximum are relative to each curve. One can see the data colapse into a single universal curve approached by a stretched exponential [5]. This is not observed for the relative $h$ index (Fig.1-B inset). This universal behavior allows comparisons among different fields.

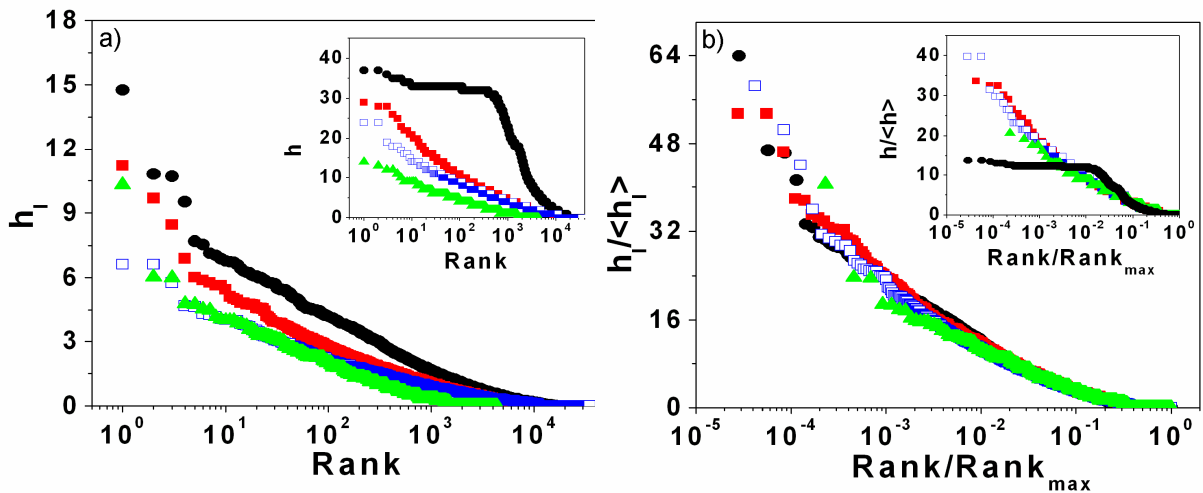

**Figure 1.** **(a)** The index $h_I$ as a function of the ranking R in four different research fields (● Physics, ■ Chemistry, □ Biology and ▲ Mathematics) in Brazil. **Inset:** The same for the $h$ index. The $h_I$ curves, in contrast to $h$ curves, have the same functional shape. **(b)** The index $h_I/\langle h_I \rangle$ as a function of the ranking $R/R_{max}$. A single universal curve is found. **Inset:** Data colapse is not obtained for $h$ curves.

# UNIVERSAL BEHAVIOR OF A RESEARCH PRODUCTIVITY INDEX
P. D. Batista, M. G. Campiteli, O. Kinouchi and A. S. Martinez
(supplementary material)

We have created the Brazilian Science Indicators (BSI) database from Thomson ISI Web of Science database. We have considered all documents published from 1945 up to 2004. The search was performed separately for each year. The choose for the Brazilian community in this work is due to the fact that the ISI Web of Science limits the searching in 100,000 papers, being thus impossible to compile a complete database for countries that have a greater annual productivity as the USA. Our database contains information about 188,909 bibliographical references which includes kind of publication, full reference, citations received yearly (up to June 2005), authors´ names and addresses, including the institutions, cities, states and country. Among all publications, we have considered only 150,323 articles, 24,164 meeting abstracts, 5,541 notes, 3,577 letters and 2,333 reviews. These documents have been classified into the following research fields: Physics, Chemistry, Biology/Biomedics and Mathematics using the *tag subject* of each document.

To further examine the robustness of our index, we have calculated $h_I$ for the top ten scientists listed by Hirsch [1] in physics and in life sciences. The results are listed in table 1. As stated by [1], the mean $h$ values for both disciplines are strongly different, as well as the maximum $h$ values, reflecting the differences between the research areas. This difference is also markedly reflected in the co-authorship patterns for both the areas. In fact, we have found a strong correlation between the rank curves for these top scientists, with $N_{Bio} \sim 2N_{Phys}$ (data not shown) and we conjecture that this correlation occurs for the rest of these communities. Thus when we normalize by the number of co-authors, the new index $h_I$ presents similar mean and maximum values as can be seen in Table 1.

| Top *h* – Physicists | | Top $h_I$ – Physicists | | Top *h* – Biologists | | Top $h_I$ – Biologists | |
|---|---|---|---|---|---|---|---|
| Witten E | 110 | Witten E | 65.05 | Snyder SH | 192 | Snyder SH | 51.92 |
| Heeger AJ | 107 | Fisher ME | 49.01 | Baltimore D | 160 | Baltimore D | 39.20 |
| Cohen ML | 104 | deGennes PG | 44.26 | Gallo RC | 154 | Moncada S | 33.66 |
| Anderson PW | 98 | Anderson PW | 38.99 | Chambon P | 153 | Dinarello CA | 26.59 |
| Gossard AC | 92 | Cohen ML | 26.44 | Vogelstein B | 152 | Chambon P | 25.52 |
| Weinberg S | 88 | Heeger AJ | 23.12 | Moncada S | 144 | Evans RM | 21.05 |
| Fisher ME | 88 | Bacall JN | 20.33 | Dinarello CA | 138 | Gallo RC | 20.56 |
| Cardona M | 87 | Gossard AC | 17.20 | Kishimoto T | 134 | Vogelstein B | 18.57 |
| deGennes PG | 79 | S Weinberg | 15.45 | Evans RM | 128 | Kishimoto T | 16.67 |
| Bacall JN | 76 | M Cardona | 6.21 | Ulrich A | 120 | Ulrich A | 16.57 |
| Mean ± sd | 92.9 ± 11.56 | | 30.6 ± 18.1 | Mean ± sd | 142.3 ± 30.0 | | 25.6 ± 13.0 |

**Table 1:** Top ten scientists listed in [1] for physics and life sciences. The *h* indexes have been computed in the ISI Web of Science querying by the authors' names as written above. For the $h_I$ we have calculated the mean number of authors in the *h* papers $<N_a>$. This new index is calculated as $h_I = h/<N_a>$. Mean and standard deviation are presented for each index.